\documentclass{article}
\usepackage[accepted]{vietnam} 
\usepackage{natbib}
\usepackage{graphicx}      
\usepackage{wrapfig}
\begin{document}
 
\twocolumn[
\title{Filamentary Structure of Orion A}
\titlerunning{Filamentary Structure of Orion A}
\author{ S\"umeyye Suri, Peter Schilke, \'Alvaro S\'anchez-Monge, and the CARMA Orion Consortium }{sumeyyesuri@gmail.com}
\address{{{\sc \textit{I}}. Phsyikalisches Institut, Universit\"at zu K\"oln, Z\"ulpicher Str.~77, 50937 K\"oln, Germany}
}
\vskip 0.5cm 
]

\begin{abstract}

{ We present filamentary structure analysis based on the CARMA Orion Project through which we obtained high resolution (0.01pc) and high dynamic range (2~deg in DEC) images of the Orion A molecular cloud. We show a preliminary emission map of  the $^{13}$CO emission with the addition of identified filaments. The data product resulted from 1500~hr of observing time at CARMA interferometer, and the Nobeyama 45~m telescope. We address basic filament properties along with calculated typical kinematical properties. Furthermore, we present a pilot study where we compare the observed [C{\sc II}] and [C{\sc I}] emission to $^{13}$CO emission towards 6 selected filaments in Orion~A. }
\end{abstract}

\section{Introduction}
Star formation takes place in the coldest ($\sim 10$~K) and densest ($10^3$--$10^6$~cm$^{-3}$) clumps of molecular clouds. Over the past years, the {\it Herschel} Space Observatory has significantly improved our understanding of the star formation process as stars and star clusters have been found to form along and at the junctions of the filaments (e.g.\ Schneider et al.\ 2012) which pervade molecular clouds on all scales observed. The observational evidence favors a model concerning the origin of filaments, where they form as a result of compression by the large-scale turbulence. They further fragment into prestellar cores when they are gravitationally unstable, i.e.\ super-critical (e.g.\ Andr\'e et al.\ 2013). Interstellar filaments that are observed with {\it Herschel} have shown a characteristic width of 0.1~pc, regardless of the corresponding column densities (e.g.\ Arzoumanian et al.\ 2011). Even though {\it Herschel} provided extensive dust continuum maps, they lacked the essential kinematic information. It is worth noting that up to date, only a small number of studies have been performed to trace the kinematics of the gas via spectral line observations, merely covering selected filaments rather than large scale maps. Nevertheless, with different spectral lines tracing various density regimes, these studies indicated that filaments may act as arteries of molecular clouds as they carry mass from the bulk of the cloud towards star forming cores (e.g.\ Schneider et al.\ 2010, Peretto et al.\ 2013).
\section{Observations}
In order to image an extensive region towards the closest host of active high-mass star formation with high resolution and high sensitivity, the CARMA Orion Key Project (PI: J.~Carpenter) was granted $\sim$ 1500 hr of observing time at the Combined Array for Millimeter-Wave Astronomy (CARMA). For the purpose of the below analysis we elaborate on the CARMA, the Nobeyama 45m, SOFIA, and APEX observations.
\subsection{CARMA Observations}
The observations were carried out using the CARMA interferometer between 2012 and 2014 using 15 antennas (6 and 10~m dishes) of the array. A 2 deg$^2$ region towards Orion A is observed using CARMA D and E arrays resulting in an angular resolution og 6~arcsec (i.e. spatial resolution of 0.012~pc). 

The spectral setup of the correlator is shown in Table~\ref{table:corr_setup}. Outflow and diffuse gas tracers (CO and $^{13}$CO), warm dense gas tracers (CS and C$^{18}$O), cold dense gas tracer (CN) and shock tracers SO along with 3 mm continuum were observed simulatenously. 

The large mosaic was divided into 228 subfields of 6 arcmin$^2$. Each individual subfield is observed in three different hour angles to be able to benefit from the Earth's rotation in filling the \textit{uv}-space and resulting in sidelobes of only $\sim$10\%.  The RMS noise in 0.5 km s$^{-1}$ channels is be 1~K for $^{12}$CO and 0.5~K for the other molecules.
\subsection{45m Observations}

The $^{13}$CO and C$^{18}$O observations  were performed during 2010 and 2013 with the 45~m telescope at the Nobeyama Radio Observatory (NRO), and are described in detail in Shimajiri et al.~(2014). A new set of CO data is being acquired at the 45~m telescope using the new receiver FOREST. At present, these data are being combined with the CARMA data in order to achieve a better sensitivity in the final products.
\subsection{APEX Observations}

In order to trace a more diffuse gas component than is observed with CARMA, we observed a selected set of filaments in [C{\sc I}] emission in APEX, and in [C{\sc II}] with SOFIA (see Section~\ref{subsection:sofi}). Observations of 6 selected filaments were completed in September 2015. 2.6 x 2.6 arcmin sized OTF maps of 6 filaments were obtained with an rms level of 0.3~K at 0.5 km~s$^{-1}$ channels. Using the FLASH460 and FLASH345 instruments we have observed $^{12}$CO(3--2) at 345.796 GHz and $^{13}$CO(3--2) 330.587 GHz along with our intended target [C{\sc I}] at 492.16 GHz.
\subsection{SOFIA Observations}\label{subsection:sofi}
Observations of 4 of the selected filaments were completed in February 2016. We utilized the upGREAT instrument onboard SOFIA to observe [C{\sc II}] emission at 1900.5 GHz. After 1.7~h of integration time we obtained a $\sim$0.8~K noise level at a channel spacing of 0.77 km~s$^{-1}$ for each filament. The angular resolution of SOFIA at the [C{\sc II}] frequency is 15.1 arcsec.

\section{Results}
\subsection{Filament Properties}
By running the filament finding algorithm DisPerSE (Sousbie et al.~2011) on the 3D (PPV) data cube, we identified $\sim$250 filaments. These filaments are shown as colored segments in Fig.~\ref{fig:carmaOrion}. We found that in contrast to the smooth filament intensity profiles in the literature, filaments in $^{13}$CO show complex substructures. Each filament is coherent within a velocity range of 1 km~s$^{-1}$. Moreover, the filament lengths vary between 0.3 -- 3.5~pc, and widths between 0.04 -- 0.25 pc. We attribute larger filament widths to either (1) existence of substructure or (2) filaments being diffuser. From the velocity fields, along with the calculated masses we find  typical accretion rates of a few times 10$^{-4}~$M$_{\odot}$~yr$^{-1}$ (Suri et al.~in prep.). These values are consistent with what is found in other filamentary clouds such as Mon-R2 (Trevino-Morales et al.~2015). 
\subsection{$^{\mathit{13}}$CO emission compared to [C{\sc \textit{I}}] and [C{\sc \textit{II}}]}
\begin{figure*}[!ht]
	\centering

    \includegraphics[scale=0.6]{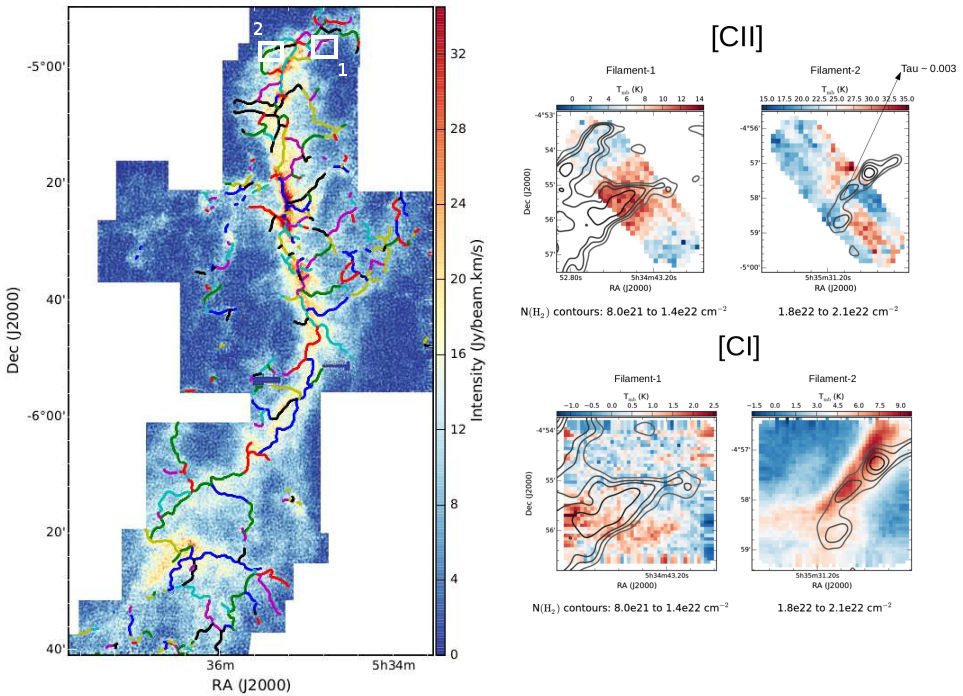}
    \caption{Property profile of the diverse library compared to the compound pool.}
    \label{fig:carmaOrion}
\end{figure*}

In Fig.~\ref{fig:carmaOrion} we also present the comparison of $^{13}$CO to that of [C{\sc I}] and [C{\sc II}] for 2 selected filaments that we further refer as Filament-1 and Filament-2. The location of Filament-1 and Filament-2 are indicated with white boxes on the left panel of Fig~\ref{fig:carmaOrion}. In the top right panel, the [C{\sc II}] single emission channel (in color), is compared to the $^{13}$CO integrated emission (in contours). The [C{\sc II}] emission towards Filament-1 peaks at 10 km~s$^{-1}$ and spacially coincides with the $^{13}$CO emission. For Filament-2, on the other hand, the [C{\sc II}] emission peaks around (above and below) the $^{13}$CO filament, indicating that it might be tracing a more diffuse component of the gas. In order to check whether there is an optically thick gas component at the ridge of Filament-2 that blocks the [C{\sc II}] emission, the optical depth of H$_2$ at the frequency of [C{\sc II}] (1900.5 GHz) is computed and found to be 0.003. Therefore, we conclude that [C{\sc II}] emission indeed peaks at the observed position, and there is not enough dust in the line-of-sight to block emission that would come from the exact position of the filament. The bottom right panel of Fig.~\ref{fig:carmaOrion} shows [C{\sc I}] emission (in color), compared to the $^{13}$CO emission in contours. While for Filament-2, the bulk of the [C{\sc I}] emission shows a similar spatial distribution, Filament-1 has a component (redshifted) that is not traced by the $^{13}$CO emission. To conclude, aforementioned findings on the correlation of $^{13}$CO and [C{\sc I}] indicates that there is an evolutionary difference between the two filaments. We speculate that Filament-1 is more evolved, therefore accreted most of the surrounding atomic gas, while Filament-2 is still surrounded by more diffuse gas that shows up in the form of atomic Carbon.
\section{Discussion}
In this section, we would like to address the fruitful audience discussion. Firstly, the $^{13}$CO radial profiles show multiple peaks throughout the map which raised the following concern, peaks might indicate existence of multiple sub-filaments, however, the significance of the peaks must be quantitatively analyzed. With the newly combined C$^{18}$O data, we will be able to perform a thorough analysis of the optical depth, and investigate this issue further. Secondly, the kinematics of the [C{\sc II}] and [C{\sc I}] were of interest. Therefore, we would like to point out a thought-provoking point to close up the discussion. For Filament-1 the [C{\sc II}] emission spacially coincides with $^{13}$CO. However, the [C{\sc II}] reaches its peak intensity is at a lower velocity than   $^{13}$CO which indicates the emission is blue-shifted and might indicate a low density ionized shell around the filament.

\begin{table*}[htbp]
  
  \caption{Correlator setup for CARMA observations.}
  \centering

  \begin{tabular}{l c c c c c} 
    \hline
     \multicolumn{1}{c}{Band}
    &\multicolumn{1}{c}{Transition}
    &\multicolumn{1}{c}{Frequency}
    &\multicolumn{1}{c}{Bandwidth}
    &\multicolumn{1}{c}{$\Delta$V}    
    \\
    \hline    
    &\multicolumn{1}{c}{}
    &\multicolumn{1}{c}{GHz}
    &\multicolumn{1}{c}{MHz}
    &\multicolumn{1}{c}{km/s}
    &\multicolumn{1}{c}{}\\
    1 & $^{12}$CO~J=1-0 & 115.271204 & 31 & 0.25 \\
    2 & CN 1-0, J= 3/2 - 1/2 & 113.490982 & 8 & 0.07 \\
    3 & $^{13}$CO J=1-0 & 110.201353 & 8 & 0.07 \\
    4 & C$^{18}$O J=1-0 & 109.782160 & 8 & 0.07 \\
    5 & SO 2(3)-1(2) & 109.252184 & 31 & 0.26 \\
    6 & CS J = 2-1 & 97.980968 & 8 & 0.08 \\
    7 & continuum & 101 \& 111.8 & 500 & -- \\
    8 & continuum & 102 \& 110.8 & 500 & --\\
    \hline  
  \end{tabular}
   \label{table:corr_setup}
\end{table*}

\section*{Acknowledgments}
The authors acknowledge funding by the Deutsche Forschungsgemeinschaft (DFG) via the Sonderforschungs-bereich SFB 956 Conditions and Impact of Star Formation (subproject A4) and the Bonn-Cologne Graduate School.



\section*{References}
Andr\'e et al.\ 2013, arXiv:1312.6232 \\
Arzoumanian et al.\ 2011, A\&A, 529, L6 \\
Peretto et al.\ 2013, A\&A, 555, A112 \\
Schneider et al.\ 2010, A\&A, 520, A4 \\
Shimajiri ~et al.\ 2014, A\&A, 564, A68 \\
Sousbie\ 2011, MNRAS, 414, 350 \\
Trevino-Morales 2015, PhD Thesis, Universidad de Granada \\








\end{document}